\documentclass[a4paper,11pt]{article}
\usepackage{pos}
\usepackage{hyperref}

\title{Astro-COLIBRI: Empowering Citizen Scientists in Multi-Messenger Astrophysics}
\ShortTitle{Astro-COLIBRI}

\author*[a]{Fabian Sch\"ussler}
\author[a]{B. Cornejo}
\author[a]{M. Costa}
\author[a]{I. Jaroschewski}
\author[a]{W. Kiendrébéogo}
\onbehalf{on behalf of the Astro-COLIBRI team}
\affiliation[a]{IRFU, CEA, Université Paris-Saclay, Gif-sur-Yvette, France}

\emailAdd{fabian.schussler@cea.fr}
\emailAdd{astro.colibri@gmail.com}

\abstract{In the era of real-time astronomy, citizen scientists play an increasingly important role in the discovery and follow-up of transient astrophysical phenomena. From local astronomical societies to global initiatives, amateur astronomers contribute valuable observational data that complement professional efforts. Astro-COLIBRI facilitates these contributions by providing a user-friendly platform that integrates real-time alerts, data visualization tools, and collaborative features to support astronomers at all levels.

The Astro-COLIBRI Citizen Science Program provides engagement opportunities across multiple scales. At the grassroots level, we collaborate with local astronomy clubs, equipping them with accessible tools for transient event monitoring. National and international networks, such as RAPAS in France, leverage Astro-COLIBRI’s real-time capabilities for coordinated observations. On a global scale, we actively participate in high-impact citizen science and capacity building initiatives, including the International Astronomical Union (IAU) Citizen Science Program and the “Open Universe” initiative led by the United Nations Office for Outer Space Affairs (UNOOSA). These collaborations enhance the accessibility of real-time astrophysical data and foster inclusive participation in cutting-edge astronomy.

In this contribution, we will present the Astro-COLIBRI Citizen Science Program, highlighting its technical framework, community impact, and case studies of successful amateur contributions. We will showcase how our platform facilitates the rapid exchange of information between professional and amateur astronomers, democratizing access to multi-messenger astrophysics and enabling the global community to contribute meaningfully to time-domain discoveries.}

\ConferenceLogo{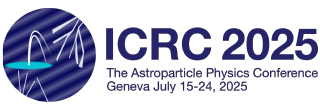}

\FullConference{39th International Cosmic Ray Conference (ICRC2025)\\
 15–24 July 2025\\
Geneva, Switzerland\\}

\begin{document}
\maketitle
\section{The New Era of Time-Domain Astronomy}
The field of astronomy is undergoing a profound transformation, driven by the rise of Time-Domain And Multi-Messenger (TDAMM) Astrophysics. Over the last decade, we have moved beyond the study of a static cosmos to an era focused on transient phenomena—brief, often cataclysmic events that provide unique insights into the most energetic processes in the Universe. These events include the explosive deaths of stars as supernovae (SNe), the enigmatic flashes of Fast Radio Bursts (FRBs), the colossal energy release of Gamma-Ray Bursts (GRBs), and the merging of compact objects like neutron stars and black holes, which ripple through spacetime as Gravitational Waves (GWs).

The scientific promise of this new era is immense. By combining observations of electromagnetic radiation across all wavelengths with "messengers" like neutrinos and GWs, we can build a complete and robust picture of these cosmic events. However, this potential is accompanied by a significant challenge: a deluge of data. Next-generation observatories and all-sky surveys, such as the Zwicky Transient Facility (ZTF) and the upcoming Vera C. Rubin Observatory, are generating an unprecedented stream of detections, numbering in the millions per night. To harness this information, discoveries must be rapidly communicated, filtered, and contextualized to enable crucial follow-up observations by a global network of observatories.

It is precisely this challenge that the Astro-COLIBRI platform was developed to address~\citep{ApJS_256_5, galaxies-11-00022}. Astro-COLIBRI serves as a comprehensive, top-level science platform that aggregates, evaluates, and contextualizes transient alerts in real time. It is designed to be a powerful tool for professional astronomers while also being intuitive and accessible, explicitly empowering the rapidly growing community of citizen scientists to participate in cutting-edge research. This proceeding details the architecture and features of the Astro-COLIBRI platform, with a special focus on its framework for fostering citizen science and Pro-Am (professional-amateur) collaborations.

\section{The Astro-COLIBRI Platform}

\begin{figure}[!t]
\centering
  \includegraphics[width=.75\linewidth]{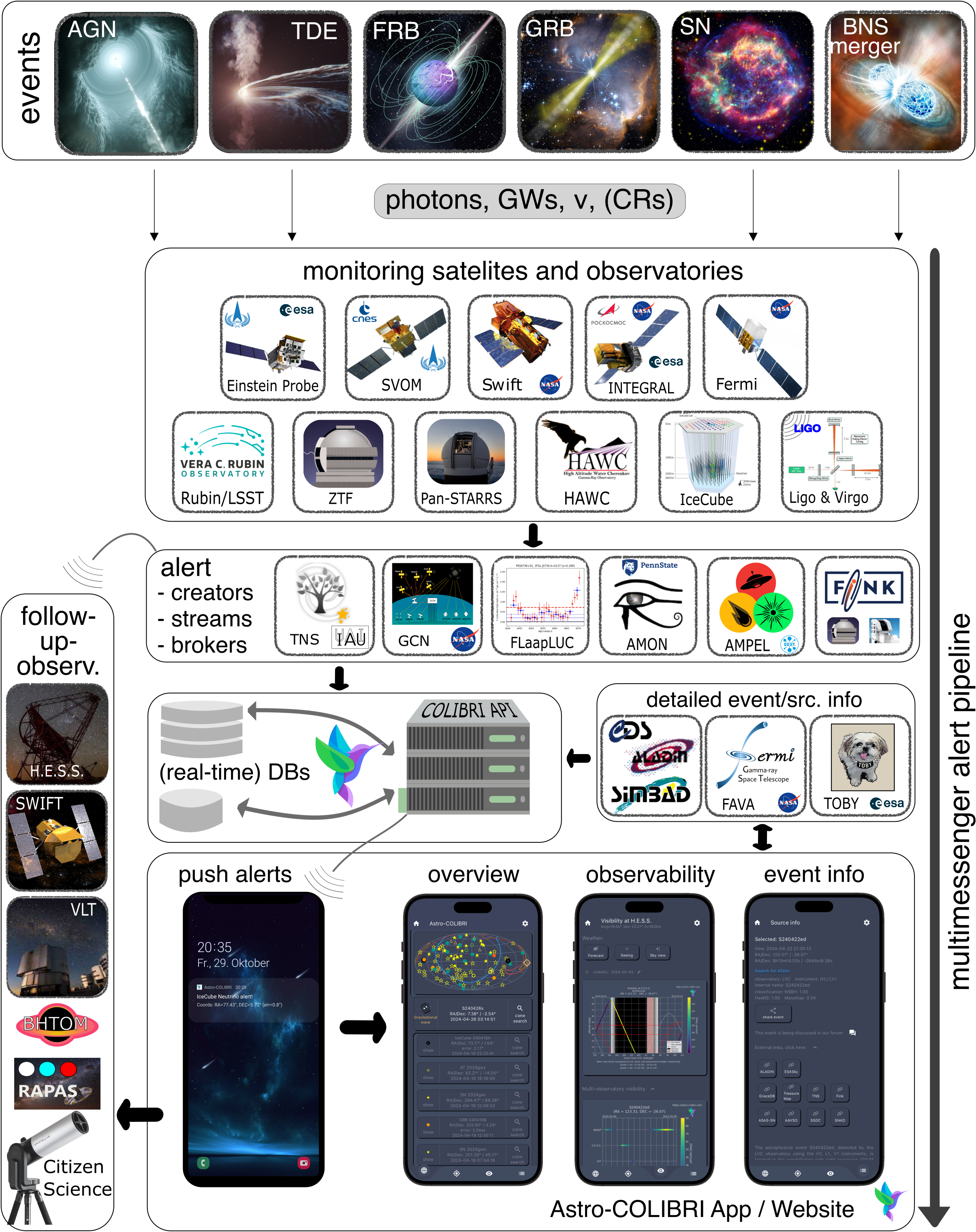}
  \caption{Overview of the real-time information flow from the cosmic events to users of the Astro-COLIBRI platform.}
  \label{fig:flow}
\end{figure}

Astro-COLIBRI is conceived as a central point in the ecosystem for real-time TDAMM studies. Its core mission is to streamline the complex process of identifying high-priority transient events and coordinating the necessary follow-up observations. This is achieved through a suite of powerful tools accessible via a web interface and dedicated mobile applications for both iOS and Android.

\subsection{System Architecture and Data Flow}
The platform is built on a modern, modular, and cloud-based architecture optimized for performance, reliability, and scalability. The main components are:
\begin{itemize}
    \item \textbf{RESTful API:} The system's backbone is a public RESTful API developed in Python using the Flask framework. It handles all core logic, data processing, and communication with external services. The API endpoints are openly accessible allowing the integration of Astro-COLIBRI functionalities in external tools and software.
    \item \textbf{Databases:} Astro-COLIBRI utilizes a dual-database system. A MongoDB instance serves as the main persistent storage for all event and source data. A Firebase Firestore real-time database is used to stream live data directly to the user clients, ensuring the interface is always up-to-date with negligible latency.
    \item \textbf{Real-Time Listeners:} Dedicated servers constantly listen to a multitude of alert streams, including VOEvents from brokers like the General Coordinates Network (GCN), notices from the Transient Name Server (TNS), and other community channels.
    \item \textbf{Cross-Platform Clients:} To ensure wide accessibility, the user interfaces—website and mobile apps—are built from a single codebase using the Flutter framework. This allows for rapid development and a consistent user experience across all devices.
    \item \textbf{Cloud-Based Alerting:} A flexible notification system using Firebase Cloud Messaging (FCM) sends real-time push alerts to users' mobile devices based on their personalized filter criteria.
    \item \textbf{Discussions:} TDAMM science is often a new and unfamiliar field for citizen scientists. To address this, the Astro-COLIBRI platform includes an open discussion forum that enables users to exchange information about recent events, share their observations, and engage directly with the development team.     
\end{itemize}
As illustrated in Fig.~\ref{fig:flow}, the data flows from monitoring observatories, through alert brokers, into the Astro-COLIBRI listeners. The API then processes each alert, stores it in the internal databases pushing it out to users in real time. 

\subsection{Core Functionalities}
Astro-COLIBRI offers a rich set of features for both professional and amateur astronomers:
\begin{itemize}
    \item \textbf{Unified Event Dashboard:} The platform displays a comprehensive list of recent transients, which can be filtered by the detecting observatory or event type (GRB, SN, GW, Neutrino, etc.), time, sky position, and many other parameters.
    \item \textbf{Interactive Sky Map:} An interactive sky map visualizes event locations, including their uncertainty regions, and overlays cataloged sources, allowing for instant spatial correlation analysis. A timeline feature further aids in visualizing temporal correlations between events.
    \item \textbf{Cone Search:} A powerful cone search functionality allows users to investigate any transient event in the context of other known sources and past events within a specified radius.
    \item \textbf{Observation Planning:} The platform provides detailed visibility plots showing when an event will be observable from a given location. Users can choose from a list of professional observatories or define their own custom sites. For events with large localization uncertainties, like those from gravitational wave detectors, Astro-COLIBRI integrates \texttt{tilepy}~\citep{SeglarArroyo2024}, a tool for generating optimized observation plans to efficiently cover the search area.
    \item \textbf{External Platforms:} A set of dedicated links provide access to a large variety of external platforms providing detailed additional information about each event.
\end{itemize}

\section{Empowering Citizen Science}
A defining feature of Astro-COLIBRI is its profound commitment to citizen science. In an era where professional telescope time is scarce, the global network of skilled amateur astronomers with increasingly sophisticated equipment represents a vital and powerful resource for TDAMM. Astro-COLIBRI is aiming to bridge the gap between professionals and amateurs, creating a truly collaborative environment. All components of the platform, including the mobile and web interfaces, API, and discussion forum are freely accessible and operate without advertisements.

Astro-COLIBRI is an official partner of Open Universe~\citep{OpenUniverse}, an initiative of the United Nations under the auspices of the UN Office for Outer Space Affairs (UNOOSA), which aims to increase the availability and accessibility of space science data. We are also an active member of the Executive Committee Working Group on Professional-Amateur Relations in Astronomy (PARC~\citep{IAU_PARC}) of the International Astronomical Union, underscoring our dedication to fostering synergies between communities at a global scale. The Astro-COLIBRI team supports time-domain observations by amateur astronomers worldwide and actively collaborates with several dedicated programs and networks. These include the french GEMINI ProAm program~\citep{Gemini} and especially the RAPAS network~\citep{RAPAS}, coordinating targeted follow-up observations of transient alerts. We also work closely with the Unistellar citizen science program on Cosmic Cataclysms, supported by the SETI Institute, which mobilizes thousands of telescope users to contribute to cutting-edge research~\citep{seti_blog}.

Beyond observational efforts, Astro-COLIBRI is deeply committed to science communication and public engagement. Through hands-on activities, interactive demonstrations, and collaborations with science festivals, we aim to share the excitement of TDAMM astrophysics with the general public—especially younger audiences. These efforts are central to our mission of promoting STEM education, inspiring the next generation of scientists, and fostering an inclusive, participatory approach to astronomical discovery.

\subsection{Features for Citizen Scientists}
While primarily aimed at the global community of (professional) astrophysicists, Astro-COLIBRI includes several features specifically tailored for the amateur astronomy community:
\begin{itemize}
    \item \textbf{Ease of Use:} The entire platform, particularly the mobile applications, is designed with a user-friendly and intuitive interface, making the vast and complex landscape of TDAMM data accessible to users of all experience levels. First-time users are greeted with an onboarding guide and will then enter a simplified interface that highlights only the most essential information. More experienced users can activate \emph{Science} mode to access the full range of data and advanced features.
    \item \textbf{Accessible Alerts:} A general "Bright Optical Transients" filter notifies users of any transient (SN, nova, etc.) brighter than magnitude 18, making them accessible to a wide range of amateur equipment. Another dedicated notification stream is available for the Unistellar citizen science program.
    \item \textbf{Custom Observatories:} Amateur astronomers are not limited to the pre-defined list of professional observatories. They can easily input their own precise geographic coordinates (or use their device's GPS) to get accurate visibility plots and observation plans tailored to their backyard or local observatory.
    \item \textbf{Community Building:} The platform fosters a sense of community through its public outreach at events like science festivals and through a dedicated discussion forum~\footnote{https://forum.astro-colibri.science} where users can share their observations and interact with both peers, professional astronomers, and with the Astro-COLIBRI team.
\end{itemize}

\subsection{Example of Community-Driven Integration: The N.I.N.A. Plugin}
Astro-COLIBRI's success is amplified by a growing ecosystem of community-developed tools that connect its alert data to a wider range of applications. One of the most impactful examples of this is the direct integration of Astro-COLIBRI into the popular astrophotography suite, N.I.N.A. (Nighttime Imaging 'N' Astronomy), through a dedicated plugin developed by the amateur astronomer Christoph Nieswand~\citep{nina}.

This powerful tool creates a seamless bridge between transient event discovery and amateur observation. Once the free plugin is installed, it provides a new window within N.I.N.A. that connects to the Astro-COLIBRI API. The plugin’s key innovation lies in its ability to translate alert information into immediate action. With a single click, an observer can select a target, and the plugin automatically populates all necessary coordinates into N.I.N.A.'s framing assistant and imaging sequencer. This high level of automation is critical for time-sensitive observations, allowing citizen scientists to pivot from notification to data collection in moments and thus contribute effectively valuable follow-up observations for new supernovae, flaring stars, and other transient phenomena.

\subsection{The Unistellar Citizen Science Program}
Another example for the Astro-COLIBRI commitment to citizen science is the partnership with the Unistellar network and the SETI Institute. The Unistellar network is a global fleet of thousands of smart, user-friendly telescopes operated by citizen scientists. Together, they form a distributed, worldwide observatory. The "Cosmic Cataclysms" program is a direct collaboration where Astro-COLIBRI acts as the alert distribution engine for the Unistellar network. The workflow is set up as follows:
\begin{enumerate}
    \item All-sky surveys like ZTF detect new, unclassified optical transients. These are ingested by alert brokers like ALeRCE. 
    \item Automatic filters developed by the Unistellar citizen science team selects promising, early supernova candidates that are bright enough to be observed by Unistellar telescopes (typically magnitude $<$ 16.3). 
    \item A dedicated "Unistellar" notification stream is available within the Astro-COLIBRI app. It announces new detections selected for the program.
    \item When a citizen scientist receives such an alert via the Astro-COLIBRI mobile app, a single tap on a dedicated link automatically commands their Unistellar telescope to point to the correct coordinates and begin capturing data.
    \item The collected observations are sent to a science server at the SETI Institute, where professional astronomers analyze the data to confirm the nature of the transient and study its lightcurve.
\end{enumerate}
This seamless integration transforms a complex workflow into a simple, engaging experience for citizen scientists, allowing them to contribute scientifically valuable data to confirm and study cosmic explosions. This partnership has already proven successful in observing supernovae and other cosmic cataclysms like novae.

\section{Scientific Use Case: The Discovery and Follow-up of SN 2025coe}
\begin{figure}[!t]
\centering
  \includegraphics[width=.55\linewidth]{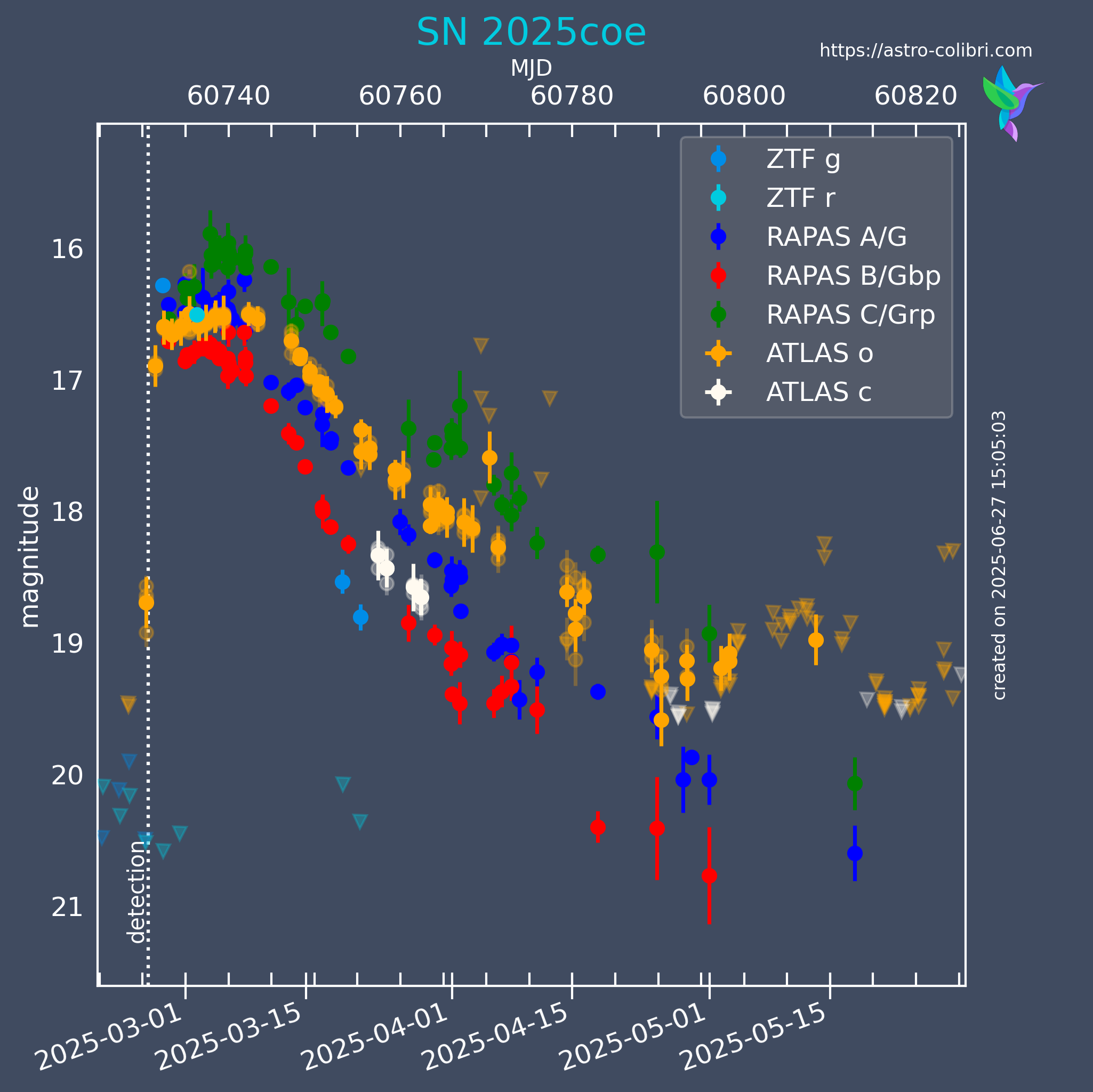}
  \caption{Example of optical lightcurves generated by Astro-COLIBRI. The event is SN 2025coe, a Ib-Ca-rich supernova detected by the amateur astronomer Koichi Itagaki~\citep{TNS2025coe}. Data from ATLAS (orange and white), and ZTF (light blue) are shown. Semi-transparent markers indicate original ATLAS data points before rebinning. Data taken by the RAPAS network of amateur astronomers~\citep{RAPAS_SN2025coe} in three filters matching the GAIA bands is shown in red, blue, and green.}
  \label{fig:SN}
\end{figure}

The discovery and subsequent detailed observation of the Type Ib Calcium-rich supernova SN 2025coe serves as a compelling example of modern Pro-Am collaboration in time-domain astronomy. Initially detected by the prolific amateur astronomer Koichi Itagaki~\citep{TNS2025coe}, the transient was reported to the Transient Name Server (TNS), and this alert was immediately ingested and relayed by the Astro-COLIBRI platform. Through its real-time mobile notifications and user-friendly interface, Astro-COLIBRI provided astronomers, both professional and amateur, with the event's critical discovery data, sky coordinates, and observability information within moments of the official report. This rapid dissemination was crucial in enabling a swift follow-up campaign by observers, including members of the French amateur network RAPAS~\citep{RAPAS_SN2025coe}. As illustrated in the lightcurve (cf. Figure~\ref{fig:SN}), the data points gathered by these citizen scientists, combined with photometry from professional surveys like ZTF and ATLAS, produced a densely-sampled, multi-band dataset essential for characterizing the supernova's evolution. For details see~\citep{Astro-COLIBRI_ICRC2025}.

\section{Conclusion and Future Outlook}
Multi-messenger astrophysics is a data-rich and time-critical field. The Astro-COLIBRI platform provides an indispensable tool for navigating this complex landscape, offering a unified, real-time view of the transient sky. It successfully integrates a vast ecosystem of alert streams, databases, and analysis tools into a single, intuitive interface for professionals and amateurs alike.

Crucially, Astro-COLIBRI is a pioneer in the democratization of astronomy. By providing tailored tools and fostering collaborations like the one with the Unistellar network, it empowers citizen scientists to move beyond outreach and make genuine, scientifically impactful contributions to cutting-edge research. The platform is continuously evolving, with plans to integrate more alert brokers, expand its feature set with tools like the "Astro-COLIBRI GPT" chatbot, and deepen its connections with the global astronomy community. Astro-COLIBRI is not just a tool for observing the cosmos, but a platform for building a more connected, collaborative, and inclusive future for science.

\section*{Acknowledgements}
The authors acknowledge support by Observatoire de Paris via the RAPAS project. We also acknowledge the support of the French Agence Nationale de la Recherche (ANR) under reference ANR-22-CE31-0012. This work was also supported by the European Union's Horizon 2020 Programme under the AHEAD2020 project (grant agreement n. 871158) and by the Horizon Europe Research and Innovation programme under the ACME project (grant agreement n. 101131928).


\begin{thebibliography}{99}
    \bibitem{ApJS_256_5} P. Reichherzer, F. Schüssler, V. Lefranc, et al., "Astro-COLIBRI—The COincidence LIBrary for Real-time Inquiry for Multimessenger Astrophysics", \href{https://iopscience.iop.org/article/10.3847/1538-4365/ac1517}{2021, ApJS, 256, 5.}
    
    \bibitem{galaxies-11-00022} P. Reichherzer, F. Schüssler, V. Lefranc, et al., "Astro-COLIBRI 2—An Advanced Platform for Real-Time Multi-Messenger Discoveries", \href{https://www.mdpi.com/2075-4434/11/1/22}{2023, Galaxies, 11, 22}
    
     \bibitem[Seglar-Arroyo et al.(2024)]{SeglarArroyo2024}
Seglar-Arroyo, M., Ashkar, H., de Bony de Lavergne, M., \& Sch\"ussler, F., "Cross Observatory Coordination with tilepy: A Novel Tool for Observations of Multimessenger Transient Events", \href{https://iopscience.iop.org/article/10.3847/1538-4365/ad5bde}{2024, ApJS, 274, 1.}



\bibitem{OpenUniverse}OpenUniverse, \url{https://openuniverse.cbpf.br}

\bibitem{IAU_PARC}IAU Executive Committee WG Professional-Amateur Relations in Astronomy, \url{https://iau.org/WG330/WG330/Home.aspx}

\bibitem{Gemini}ProAm Gemini, \url{https://gemini.obspm.fr/}

\bibitem{RAPAS}RAPAS, \url{https://rapas.imcce.fr/}
    
    \bibitem{seti_blog} SETI Institute, "SETI Institute \& Unistellar Collaborate with Astro-COLIBRI to Observe Cataclysmic Events", \url{https://www.seti.org/news-archive/news-archive-detail/?id=5464}

    \bibitem{nina}C. Niewand, "Astro-COLIBRI N.I.N.A. Plugin",\\ \url{https://github.com/chrisastrophoto/Nina.AstroColibri}
    
    \bibitem{TNS2025coe} K. Itagaki, TNS Astronomical Transient Report No. 245772\\ \url{https://www.wis-tns.org/object/2025coe/discovery-cert}

    \bibitem{RAPAS_SN2025coe} T. Midavaine et al. (RAPAS), TNS AstroNote 2025-200, \url{https://www.wis-tns.org/astronotes/astronote/2025-200}

    \bibitem{Astro-COLIBRI_ICRC2025} F. Sch\"ussler et al., "Astro-COLIBRI: A Comprehensive Platform for Real-Time Multi-Messenger Astrophysics", Proceedings ICRC2025, \href{https://pos.sissa.it/501/966}{PoS(ICRC2025)966}
    
\end{thebibliography}
\end{document}